# Bayesian Mechanism Design for Budget-Constrained Agents

Shuchi Chawla[*]    David Malec[†]    Azarakhsh Malekian[‡]

**Abstract**

We study Bayesian mechanism design problems in settings where agents have budgets. Specifically, an agent's utility for an outcome is given by his value for the outcome minus any payment he makes to the mechanism, as long as the payment is below his budget, and is negative infinity otherwise. This discontinuity in the utility function presents a significant challenge in the design of good mechanisms, and classical "unconstrained" mechanisms fail to work in settings with budgets. The goal of this paper is to develop general reductions from budget-constrained Bayesian MD to unconstrained Bayesian MD with small loss in performance. We consider this question in the context of the two most well-studied objectives in mechanism design—social welfare and revenue—and present constant factor approximations in a number of settings. Some of our results extend to settings where budgets are private and agents need to be incentivized to reveal them truthfully.

[*]Computer Sciences Dept., University of Wisconsin - Madison. `shuchi@cs.wisc.edu`.
[†]Computer Sciences Dept., University of Wisconsin - Madison. `dmalec@cs.wisc.edu`.
[‡]EECS, Northwestern University. `a-malekian@northwestern.edu`.

# 1  Introduction

Auction and mechanism design have for the most part focused on agents with quasilinear utility functions: each agent is described by a function that assigns values to possible outcomes, and the agent's utility from an outcome is her value minus any payment that she makes to the mechanism. This implies, for example, that an agent offered an outcome at a price below her value for the outcome should in the absence of better alternatives immediately accept that outcome. This simple model fails to capture a basic practical issue—agents may not necessarily be able to afford outcomes that they value highly. For example, most people would value a large precious stone such as the Kohinoor diamond at several millions of dollars (for its resale value, if not for personal reasons), but few can afford to pay even a fraction of that amount. Many real-world mechanism design scenarios involve financially constrained agents and values alone fail to capture agents' preferences. Budget constraints have frequently been observed in FCC spectrum auctions [5, 7], Google's auction for TV ads [18], and sponsored search auctions, to take a few examples.

From a theoretical viewpoint, the introduction of budget constraints presents a challenge in mechanism design because they make the utility of an agent nonlinear and discontinuous as a function of the agent's payment—the utility decreases linearly with payment while payment stays below the budget, but drops to negative infinity when the payment crosses the budget. The assumption of linearity in payments (i.e. quasilinearity of utility) underlies much of the theoretical framework for mechanism design. Consequently, standard mechanisms such as the VCG mechanism can no longer be employed in settings involving budgets.

The goal of this paper is to develop connections between budget-constrained mechanism design and the well-developed theory of unconstrained mechanism design. Specifically we ask "when can budget-constrained mechanism design be reduced to unconstrained mechanism design with some small loss in performance?" We consider this question in the context of the two most well-studied objectives in mechanism design—social welfare and revenue. Some of our results assume that the mechanism knows the budgets of the agents, but others hold even when budgets are private and agents need to be incentivized to reveal them truthfully.

Recent work in computer science has begun exploring a theory of mechanism design for budget-constrained agents (see, for example, [1, 4, 10, 9, 2]). Most of this work has focused on *prior-free* or worst-case settings, where the mechanism designer has no information about agents' preferences. Unsurprisingly, the mechanism designer has very little power in such settings, and numerous impossibility results hold. For example, in the worst-case setting no truthful mechanism can obtain a non-trivial approximation to social welfare [4]. The goal of achieving good social welfare has therefore been abandoned in favor of weaker notions such as Pareto optimality [10]. For the revenue objective while approximations can be achieved in simple enough settings, e.g. multi-unit auctions [4], hardness results hold for more general feasibility constraints even in the absence of budgets. In this paper, we sidestep these impossibility results by considering Bayesian settings where the mechanism designer has prior information about the distributions from which agents' private values and private budgets are drawn.

We restrict our attention to direct revelation truthful mechanisms. Our mechanisms are allowed to randomize, and agents' utilities are computed in expectation over the randomness used by the mechanism. As is standard, we assume that both the mechanism and the agents possess a common prior from which values are drawn. While we optimize over the class of Bayesian incentive compatible (BIC) mechanisms, all of the mechanisms we develop are dominant strategy incentive compatible (DSIC) (see, for example, [17] for definitions of these solution concepts).

In addition, we require that our mechanisms satisfy the ex-post individual rationality (EPIR) constraint, namely that the payment of any agent never exceeds her value for the mechanism's outcome. This implies, in particular, that the mechanism cannot charge any agent to whom no item or service is allocated. In contrast, most previous work has enforced the individual rationality constraint only in expectation over the mechanism's randomness as well as the randomness in other agents' values (i.e. interim IR).

It is worth noting here that the EPIR constraint is not without loss in performance. Consider the following example: suppose we are selling a single item to one of $n$ agents, each with a value of $v$ with probability $1$ (that is publicly known) and a public budget of $v/n$ with probability $1$. Now, under the IIR constraint, the optimal auction asks agents to pay what they bid and offers each agent that pays at least $v/n$ a fair chance at winning the item. Each agent pays $v/n$, the item is allocated to a random agent, and the mechanism's revenue is $v$. Under the EPIR constraint, however, a mechanism can only charge the agent that wins the item and can charge this agent no more than $v/n$. As we can see, the revenue gap between the optimal IIR and the optimal EPIR mechanism gets larger and larger as $n$ grows.

It is well known that over the class of BIC IIR mechanisms, the revenue-optimal as well as welfare-optimal mechanisms are both so-called "all-pay" auctions [15, 19]. In all-pay auctions agents pay the mechanism a certain (distribution dependent) function of their value regardless of the allocation that the mechanism makes. The optimality of all-pay auctions follows by noting that any allocation rule that admits some BIC budget-feasible payment function can be implemented with an all-pay payment rule with worst-case payments that are no larger than those in any other truthful payment rule and are therefore budget-feasible. Unfortunately all-pay auctions have many undesirable properties. In many settings it is simply not feasible to force the agents to pay upfront without knowing the outcome of the mechanism. Moreover all-pay auctions may admit many Bayes-Nash equilibria (BNE), truthtelling being merely one of them. Then the fact that a certain objective is achieved when all the agents report their true types does not necessarily imply that the objective will be achieved in practice if a different BNE gets played out. Therefore, in a departure from previous work, we choose to enforce ex-post individual rationality.

## Our results and techniques

We begin our investigation with the revenue objective and give an exact characterization of the optimal mechanism for a single agent with a public budget. While in the absence of budgets the optimal mechanism is a fixed sale price and therefore deterministic, with budgets the optimal mechanism may need to randomize and offer multiple buying options to the agent. This complicates the design of optimal mechanisms in more general settings involving multiple agents or private budgets. We therefore consider approximations. When budgets are known publicly, we obtain constant factor approximations in nearly all settings where constant factor approximations are known for unconstrained mechanism design. This includes, for example, all single-parameter settings with a downwards closed feasibility constraint, but also multi-parameter settings with "unit-demand" agents and a matroid feasibility constraint (see, e.g., [6]). Our mechanisms are for the most part direct reductions to unconstrained settings, and are extremely simple.

For private budgets, the problem becomes much harder and we focus on settings with single-dimensional values. We design a novel mechanism based on "lotteries" that obtains a good approximation whenever each agent's value distribution satisfies the monotone hazard rate (MHR) condition (see Section 2 for a definition). Our mechanism's novelty lies in offering each agent a carefully constructed set of different buying options such that the best option for the agent is to either spend his entire budget or a fraction of the monopoly price for that agent. The MHR assumption is a frequently used assumption in mechanism design literature and many natural distributions satisfy it. In fact the mechanism obtains a good approximation more generally under mild technical conditions on the values and budgets. We believe that our techniques should extend to provide good approximations for arbitrary distributions.

Next we examine the welfare objective. While for revenue, the budget of an agent is a natural upper bound on the contribution of that agent to the revenue and allows us to "cap" values at the budget, for welfare this doesn't work. In fact, a mechanism can generate a non-trivial guarantee on welfare even when budgets are $0$. Consider a setting with two unit demand buyers and two items. Consider the following mechanism: the mechanism asks each agent to give a preference list of the items. If the top choices of the buyers are different, then each buyer gets allocated his top choice and welfare is maximized. Otherwise, the mechanism ignores the preferences of the buyers and computes the allocation that maximizes the social welfare ex-ante. Note that this mechanism is truthful. When agents' values for the items are i.i.d., the obtained social welfare from this example is at least $3/4$ of the maximum social welfare we can obtain with no budget constraint. On the other hand, a mechanism that "ignores" values above the budget (i.e. does not distinguish between them in the allocation function) cannot obtain an approximation better than $1/2$. The gap between the two mechanisms increases as the number of agents grows.

We again focus on single-parameter settings and public budgets, but with arbitrary downwards closed feasibility constraints. For these settings, we show a tradeoff between an approximation on budget and an approximation on welfare: for any $\epsilon$, we can get a $1/\epsilon$ approximation to welfare with respect to the welfare that an optimal mechanism can get when budgets are scaled down by a factor of $1 - \epsilon$. This mechanism has an extremely simple form: it replaces every value larger than its budget by its expectation conditioned on being larger than the budget, and runs the VCG mechanism on these modified values. Moreover, if we are willing to sacrifice EPIR in favor of the less restrictive IIR, we can convert this mechanism into a 4-approximate IIR mechanism (with no approximation on budgets).

Finally, if the value distributions satisfy the MHR condition, we achieve a $2(1 + e)$-approximation to welfare via an EPIR mechanism by reducing budget-feasible welfare maximization to budget-feasible revenue maximization.

One nice property of our reductions from budget feasible mechanism design to unconstrained mechanism design is that they are for the most part oblivious to the feasibility constraint imposed on the mechanism. They therefore work for a broad range of feasibility constraints and add minimal complexity to the mechanism design problem.

**Related work**

Several works in economics have studied characterizations of optimal BIC IIR budget-feasible mechanisms (e.g., [19, 14, 8, 15]). However, these works are generally weak in the kinds of settings they consider (typically just single-item auctions) and the kinds of value distributions they allow[1]. Laffont and Robert [14] considered single item settings where bidders have a private value and public common budget. Che and Gale [8] considered the setting with a single item and a single buyer, but allowed both the value and the budget to be private. Pai and Vohra [19] gave a more general result in which they designed an optimal auction for a single item and multiple buyers with private i.i.d. values and private budgets.

Bhattacharya et al. [3] were the first to study settings beyond single-item auctions and focused on revenue maximization. They considered a setting with heterogeneous items and additive values, and exhibited a (large) constant factor DSIC approximation mechanism as well as an all-pay auction which admits truthtelling as a BNE and in that BNE obtains a $4$-approximation. However, these results required the value distributions to satisfy the MHR condition. The mechanisms are LP-based. In contrast most of our mechanisms are easy to compute, work for general distributions, enforce EPIR, and achieve small approximation factors.

In prior-free settings few results are known for revenue maximization. Borgs et al. [4] looked at multi unit auctions for homogeneous goods where agents have private values and budgets and considered worst case competitive ratio (see also [1]). They designed a mechanism based on random sampling that maximizes revenue when the number of bidders is large.

Social welfare maximization has also been considered under budget constraints. Maskin [15] considered the setting of a single item and multiple buyers with public budgets. He defined and showed how to compute the constrained efficient mechanism, the truthful feasible mechanism under budget constraints that maximizes the expected social welfare (however, the result holds only for some distribution functions [19]). In prior-free settings for multi unit homogeneous items, Nisan et al. [10] studied Pareto efficient DSIC mechanisms with budget constraints. They showed that if the budgets are private there is no Pareto optimal incentive compatible mechanism; for public budgets they showed that there exists a unique mechanism based on the *clinching auction*. Chen et al. [9] considered a setting with multiple goods and unit demand buyers and showed how to compute competitive prices that enforce truthfulness under budget constraints if such prices exist. Finally, the work of Alaei et al. [2] stands out in their study of "soft" budgets constraints, where buyers pay an increasing interest rate for payments made above their budgets. They showed how to exactly compute the smallest competitive prices in this setting that result in an incentive compatible mechanism with an outcome in the core.

## 2  Notation and definitions

In this work, we consider instances of the Bayesian mechanism design problem where agents have single- and multi-dimensional types; instances are of the form $\mathcal{I} = (\mathbf{F}, \mathcal{S}, \mathbf{B})$.

In the single-dimensional case, $\mathbf{F} = \prod_i F_i$ is a product distribution; each agent $i$ has a single value $v_i \sim F_i$ for receiving service (and derives value $0$ if not served), and an upper limit $B_i$ on how much he or she can pay for service; and $\mathcal{S}$ is a feasibility constraint specifying which sets of agents may be simultaneously served.

In the multi-dimensional case, the seller offers a number of services indexed by $j$ to agents, and $\mathbf{F} = \prod_{i,j} F_{ij}$ is again a product distribution; each agent $i$ has a value for receiving service $j$ of $v_{ij} \sim F_{ij}$ and is interested in receiving at most one service; and the agent has a budget limit of $B_i$. Here, $\mathcal{S}$ is a feasibility constraint over pairs $(i, j)$.

We also consider settings with private budgets; in that case, we replace $\mathbf{B}$ with a distribution $\mathbf{G} = \prod_i G_i$ and agent $i$ has a budget $B_i \sim G_i$.

We focus on incentive compatible (IC) and individually rational (IR) mechanisms, and further distinguish between Bayesian IC and dominant strategy IC, and interim IR and ex post IR. See [17] for definitions of these concepts.

---

[1] E.g., [19] and [15] make the assumption that value distributions have a monotone hazard rate as well as a nondecreasing density function, unnatural conditions that few distributions satisfy simultaneously.

Let $M$ be a mechanism for the instance $\mathcal{I}$. We shall denote its expected allocation to each agent by the vector $\mathbf{x}(\mathbf{v}, \mathbf{B})$, and agents' expected payments by $\mathbf{p}(\mathbf{v}, \mathbf{B})$ (we omit the second parameter when $\mathbf{B}$ is fixed). Then the expected revenue of $M$ is $\mathcal{R}^M = \mathrm{E}_{\mathbf{v}, \mathbf{B}}[\mathbf{p}(\mathbf{v}, \mathbf{B}) \cdot \mathbf{x}(\mathbf{v}, \mathbf{B})]$, and its expected social welfare is $\mathrm{E}_{\mathbf{v}, \mathbf{B}}[\mathbf{v} \cdot \mathbf{x}(\mathbf{v}, \mathbf{B})]$

Given that an agent $i$ has a value $v$ for receiving service, and a budget constraint $B$, his or her utility from receiving the service with probability $x$ at a price $p$ is $u(v, B) = x \cdot v - p$ if $p \leq B$, and $u(v, B) = -\infty$ otherwise. A mechanism is budget feasible if it never requests an agent to make a payment above his budget.

**Virtual values and the monotone hazard rate condition**

In the absence of budget constraints, for revenue maximization Myerson in his seminal work [16] gives a characterization of the optimal mechanism as a "virtual value maximizer". Specifically, given any distribution function $F$ with density $f$, Myerson defines a *virtual value* function as follows:

$$\phi(v) = v - \frac{1 - F(v)}{f(v)}$$

We use the following characterization by Myerson of the expected revenue of BIC mechanisms in terms of their virtual surplus.

**Lemma 1.** *Consider any BIC mechanism with allocation function $\mathbf{x}$ for a single-parameter problem $\mathcal{I} = (\mathbf{F}, \mathcal{S})$. Then the expected revenue of the mechanism is exactly $\mathrm{E}_{\mathbf{v} \sim \mathbf{F}}[\sum_i x_i(\mathbf{v}) \phi_i(\mathbf{v})]$.*

A distribution is said to be *regular* if $\phi(v)$ is a non-decreasing function of $v$. When value distributions are regular, a mechanism that allocates to the feasible set that maximizes the total virtual value is BIC and optimal. For a single agent, this mechanism allocates to the agent as long as his value is above the threshold $\phi^{-1}(0)$; we call this threshold the *monopoly price* corresponding to the value distribution.

When value distributions are not regular, Myerson gives an "ironing" procedure that converts a virtual value function into an ironed virtual value function, $\bar{\phi}$, such that maximizing ironed virtual surplus results in a BIC optimal mechanism. We omit the details of the ironing.

Some of our results require a stronger condition on distributions called the monotone hazard rate condition, a common assumption in mechanism design literature. This condition is satisfied by many common distributions such as the uniform, Gaussian, exponential, and power law distributions.

**Definition 1.** *A distribution $F$ with density $f$ is said to have a **monotone hazard rate** if the function $h(v) = f(v)/(1 - F(v))$ is non-decreasing in $v$. Distributions satisfying MHR are regular.*

## 3 Maximizing revenue

We first consider the revenue objective, and begin by characterizing the optimal budget feasible mechanism for a single agent setting. The characterization relies on describing the mechanism as a collection of so-called lotteries or randomized pricings. We then consider settings with public budgets. Our general approach towards budget-constrained mechanism design in these settings is to approximate the optimal revenue in two parts: the contribution to optimal revenue by agents whose budget is binding (i.e. their budget is less than their value), and the contribution by agents whose budget is not binding (i.e. their budget is above their value). We present different mechanisms for approximating these two benchmarks. We demonstrate this approach first in the simple setting of single-parameter agents with public budgets and an arbitrary downwards closed feasibility constraint. Then in subsequent sections we extend the approach to settings involving more complicated incentive constraints—multi-dimensional values and private budgets. In private budget settings, instead of asking agents to reveal budgets directly, our mechanism once again relies on collections of lotteries to motivate agents to pay a good fraction of their budgets when their values are high enough.

## 3.1 Single agent settings with public budgets

Before presenting our general approach, we first consider the most basic version of this problem—namely a setting with one single-parameter agent and a public budget constraint. Even this simple setting, however, reveals the challenges budget constraints introduce to the problem of mechanism design. Without the budget constraint, the optimal mechanism is to offer the item at a fixed price. With budgets, however, the following example shows that a single fixed price can be a factor of 2 from optimal. After the example we proceed to characterize the optimal mechanism.

**Example 1.** *Fix $n > 1$. Consider an agent whose value for receiving an item is $v = 1$ with probability $1 - 1/n$, and is $v = n^2$ with probability $1/n$. Let the agent have a budget of $B = n$. Any single fixed price that respects the budget in this setting receives a revenue of at most $1$.*

*We now describe the optimal mechanism. The mechanism offers two options to the agent: either buy the item at price $n$, or receive the item with a probability of $n/(n+1)$ at a price of $n/(n+1)$. This generates an expected revenue of $2n/(n+1) = 2 - o(1)$.*

The optimal mechanism in the above example is what we call a *lottery menu* mechanism. A lottery is a pair $(x, p)$ and offers to the agent at a price $p$ a probability $x$ of winning. A lottery menu is a collection of lotteries that an agent is free to choose from in order to maximize his expected utility. We will now show that for any single agent setting with a public budget, the optimal mechanism is a lottery menu mechanism with at most two options.

Consider a setting $\mathcal{I}$ with a single agent with private value $v \sim F$ and a public budget $B$. Let $\phi$ be the virtual value function corresponding to $F$. For ease of exposition, throughout the following discussion we will assume that $F$ is regular and $\phi$ is non-decreasing; when $F$ is non-regular, we can merely replace $\phi$ by $\bar{\phi}$, the ironed virtual value, in the following discussion.

We first note that if $B \geq \phi^{-1}(0)$ then the unconstrained optimal mechanism is already budget feasible. Therefore, for the rest of this section we assume that $B < \phi^{-1}(0)$. Following Lemma 1, our goal is to solve the following optimization problem.

$$\max_x \int x(v)\phi(v)f(v)dv \quad \text{subject to}$$
$$\int (x_{\max} - x(v))dv \leq B \cdot x_{\max}, \quad \text{and,}$$
$$x(v) \text{ is a non-decreasing function.}$$

Here $x_{\max} \leq 1$ is the probability of allocation at the upper end of the support of the value distribution. The first constraint encodes the budget constraint. In particular, the left hand side of the inequality is the expected payment made by the agent at his highest value; the right side is an upper bound on the expected payment under EPIR because the agent can pay a maximum of $B$ when he gets allocated, and $0$ otherwise.

Let $x^*$ be the optimal solution to the above optimization problem. We make the following observations (proofs are given in Appendix B). In the following, we denote the inverse virtual value of $0$ as $v^* = \phi^{-1}(0)$ to simplify notation.

**Claim 1.** *Without loss of generality, we may assume $x^*_{max} = 1$.*

**Claim 2.** *Without loss of generality, we may assume that for all $v \geq v^*$, $x^*(v) = 1$.*

Following these claims, our optimization problem changes to the following (the monotonicity constraint on $x$ is implicit).

$$\max_x \int x(v)\phi(v)f(v)dv \quad \text{subject to}$$
$$\int (1 - x(v))dv \leq B$$
$$x(v) = 1 \quad \forall v \geq v^*$$

This can be simplified to:

$$\min_x \int_0^{v^*} x(v)(-\phi(v)f(v))dv \text{ subject to}$$

$$\int_0^{v^*} (1-x(v))dv = B$$

Note that we replace the inequality in the budget constraint with an equality. This is because if the constraint is not tight, we can feasibly reduce $x(v)$ and thereby reduce the objective function value. For the sake of brevity, we define $B' = v^* - B$, and $g(v) = -\phi(v)f(v)$. The budget constraint then changes to $\int_0^{v^*} x(v)dv = B'$. Note that $v^* \geq B' \geq 0$, and $g$ is nonnegative on $[0, v^*]$. Finally, we define the set of allocations

$$\mathcal{A} = \left\{ \begin{array}{l} \text{increasing } x : [0, v^*] \to [0, 1] \\ \text{such that } \int_0^{v^*} x(v)dv = B' \end{array} \right\}$$

Then, we can express our objective as

$$\min_{x \in \mathcal{A}} \int_0^{v^*} x(v)g(v)dv.$$

If $g$ is non-increasing on $[0, v^*]$, then we immediately have that the optimal solution is to set $x(v) = 1$ if $v \geq v^* - B' (= B)$ and 0 otherwise.

If $g$ is not non-increasing, we "iron" the function $g$ to produce a non-increasing function $\hat{g}$ with the property that any non-decreasing function $x$ that is constant over intervals where $\hat{g}$ is constant has the same integral with respect to $\hat{g}$ as with respect to $g$. Let $\widetilde{\mathcal{A}}$ be the subset of $\mathcal{A}$ containing all functions $x$ that are constant over intervals where $\hat{g}$ is constant. We obtain the following lemma. (The details of the ironing procedure and the proof of the following lemma can be found in Appendix B.)

**Lemma 2.** *For all $x \in \mathcal{A}$, there exists a $\tilde{x} \in \widetilde{\mathcal{A}}$, such that $\int_0^{v^*} x(v)g(v)dv \geq \int_0^{v^*} \tilde{x}(v)g(v)dv$.*

The lemma lets us confine our optimization to the set $\widetilde{\mathcal{A}}$:

$$\min_{x \in \mathcal{A}} \int_0^{v^*} x(v)g(v)dv = \min_{x \in \widetilde{\mathcal{A}}} \int_0^{v^*} x(v)g(v)dv$$

Finally, we define $\mathcal{A}^*$ to be a subset of $\widetilde{\mathcal{A}}$ in which functions $x$ take on at most three different values – 0, 1, and an intermediate value. The final part of our proof is to show that the optimal solution lies in this set.

**Theorem 3.** *For any single agent setting $\mathcal{I} = (F, B)$, there is an optimal mechanism with allocation rule in the set $\mathcal{A}^*$.*

*Proof.* Recall that the optimal solution $x^*$ lies in the set $\widetilde{\mathcal{A}}$. Suppose for contradiction that this function takes on two different intermediate values, $x^*(v_1) = y$ and $x^*(v_2) = z$, between 0 and 1 with $y < z$. Then, since $\hat{g}$ is non-increasing and $x^*$ is non-decreasing, we must have $\hat{g}(v_1) > \hat{g}(v_2)$. Now we can improve our objective function value by increasing $x^*$ between $v_2$ and the value at which it becomes 1, and decreasing $x^*$ between the value at which it becomes strictly positive and $v_1$, while maintaining the budget constraint. This contradicts the optimality of $x^*$. □

## 3.2 Single parameter setting with public budgets

We now consider single parameter settings with multiple agents. Let $\mathcal{I} = (\mathbf{F}, \mathcal{S}, \mathbf{B})$ be an instance of single-parameter budget-constrained revenue maximization. Define the truncated distributions $\widehat{F}_i$ as follows.

$$\widehat{F}_i(v) = \begin{cases} F_i(v) & \text{if } v < B_i; \text{ and} \\ 1 & \text{if } v \geq B_i. \end{cases} \quad (1)$$

Let $\widehat{\mathcal{I}} = (\widehat{\mathbf{F}}, \mathcal{S})$ be the modified setting where we replace $\mathbf{F}$ with $\widehat{\mathbf{F}}$ — note that for each $i$, the support of $\widehat{F}_i$ ends at or before $B_i$, and so we may remove the budgets since they place no constraint on the instance $\widehat{\mathcal{I}}$.

A mechanism for $\widehat{\mathcal{I}}$ naturally extends to $\mathcal{I}$, while satisfying budget feasibility and obtaining the same revenue. Our general technique will be to relate the revenue of a mechanism for $\mathcal{I}$ to that of a mechanism for $\widehat{\mathcal{I}}$. In general, the latter can be quite small, and so we introduce the following quantity to bound this loss. Define the set $\mathcal{B}$ as

$$\mathcal{B} = \underset{S \in \mathcal{S}}{\operatorname{argmax}} \left\{ \sum_{i \in S} B_i \,\middle|\, \forall i \in S, v_i \geq B_i \right\}. \tag{2}$$

Our basic approach is to design a BIC mechanism $\widehat{M}$ for the setting $\widehat{\mathcal{I}}$ based on the original mechanism $M$ such that we have

$$\mathcal{R}^M \leq \mathcal{R}^{\widehat{M}} + \mathrm{E}\left[\sum_{i \in \mathcal{B}} B_i\right]. \tag{3}$$

Then, the first term on the right is bounded above by the revenue of the optimal mechanism for $\widehat{\mathcal{I}}$. We further demonstrate in each case that we can bound the expectation $\mathrm{E}\left[\sum_{i \in \mathcal{B}} B_i\right]$ by another mechanism for $\widehat{\mathcal{I}}$.

We define the mechanism $\widehat{M}$ in terms of its expected allocation and payment. Let $\mathbf{x}(\mathbf{v})$ and $\mathbf{p}(\mathbf{v})$ be the expected allocations and expected payments for $M$, respectively. Define the expected allocation and expected payment rules for $\widehat{M}$ as follows. For each agent $i$ in the setting $\widehat{\mathcal{I}}$ with valuation $\hat{v}_i$, draw a corresponding $v_i$ consistent with $\hat{v}_i = \min(v_i, B_i)$; in this case that simply means $v_i = \hat{v}_i$ if $\hat{v}_i < B_i$, and $v_i \sim F_i(v \mid v \geq B_i)$ otherwise. Then $\widehat{M}$'s expected allocation and payment are given by

$$\hat{x}_i(\hat{\mathbf{v}}) = x_i(\mathbf{v}_{-i}, \hat{v}_i); \text{ and} \qquad \hat{p}_i(\hat{\mathbf{v}}) = p_i(\mathbf{v}_{-i}, \hat{v}_i),$$

respectively.

**Lemma 4.** *$\widehat{M}$ is a feasible BIC mechanism for $\widehat{\mathcal{I}}$.*

*Proof.* We first note that from the point of view of a single agent $i$, the expected allocation and price function of $\widehat{M}$ behave as though other agents' values are the same as before. Therefore, the expected allocation is still an increasing function of value and the payments satisfy BIC. We will now argue that the expected allocation function can be implemented in a way that the resulting outcome is a randomization over feasible outcomes. To do so, we first compute $x_i(\mathbf{v})$, as well as $\hat{x}_i(\hat{\mathbf{v}})$ for all $i$. Starting with the allocation returned by $\mathbf{x}(\mathbf{v})$, for every agent $i$ in this allocation, with probability $\hat{x}_i(\hat{\mathbf{v}})/x_i(\mathbf{v})$, we serve this agent, and with the remaining probability we remove her from the allocated set. Since $\mathcal{S}$ is a downward closed feasibility constraint, feasibility is maintained, and we achieve the target allocation probabilities. We remark here that our goal is to merely exhibit that $\widehat{M}$ is feasible and not to actually compute it. □

We now prove the bound (3) on $\mathcal{R}^M$.

**Lemma 5.** *Given any mechanism $M$ for $\mathcal{I} = (\mathbf{F}, \mathcal{S}, \mathbf{B})$, where $\mathcal{S}$ is downward-closed, if we define the mechanism $\widehat{M}$ for $\widehat{\mathcal{I}}$ as above, then (3) holds.*

*Proof.* In order to prove the statement, we couple the values $\mathbf{v}$ that $\widehat{M}$ draws for each $\hat{\mathbf{v}}$ with the $\mathbf{v}$ in the other expectations. So fix some corresponding pair of value vectors $\mathbf{v}$ and $\hat{\mathbf{v}}$; consider the contribution of each agent $i$ to the revenue of $M$. Split the agents into two sets $L$ and $H$, defined by

$$L = \{i | v_i \leq B_i\}; \text{ and} \qquad H = \{i | v_i \geq B_i\}.$$

Recall that for all $i \in L$, we have that $v_i = \hat{v}_i$, and so $\hat{p}_i(\hat{\mathbf{v}}) = p_i(\mathbf{v}_{-i}, \hat{v}) = p_i(\mathbf{v})$. Furthermore, since $M$ faces the downward-closed feasibility constraint $\mathcal{S}$, any subset of $H$ that $M$ serves is one of the sets $\mathcal{B}$ maximizes over. Since $M$ can never charge any agent more than their budget, we can see that

$$\mathcal{R}^M(\mathbf{v}) = \sum_{i \in L} \mathcal{R}^M_i(\mathbf{v}) + \sum_{i \in H} \mathcal{R}^M_i(\mathbf{v}) \leq \sum_{i \in L} \mathcal{R}^{\widehat{M}}_i(\hat{\mathbf{v}}) + \sum_{i \in \mathcal{B}} B_i \leq \mathcal{R}^{\widehat{M}}(\hat{\mathbf{v}}) + \sum_{i \in \mathcal{B}} B_i.$$

Taking expectations on both sides (according to the previously mentioned coupling) proves the claim. □

Note that $\mathcal{R}^{\widehat{M}}$ can be easily achieved by simply running the (unconstrained) revenue-optimal mechanism over $\widehat{\mathcal{I}}$. It remains to be shown that we can, in fact, upper bound $\mathrm{E}[\sum_{i \in \mathcal{B}} B_i]$ also by the revenue of the same (unconstrained) revenue-optimal mechanism over $\widehat{\mathcal{I}}$.

**Lemma 6.** *There exists a mechanism $M_\mathcal{B}$ for the setting $\widehat{\mathcal{I}}$ such that $\mathrm{E}_{\mathbf{v} \sim \mathbf{F}} \left[ \sum_{i \in \mathcal{B}} B_i \right] \leq \mathcal{R}^{M_\mathcal{B}}$.*

*Proof.* We define the mechanism $M_\mathcal{B}$ as implementing the allocation rule $\mathcal{B}$. Note that membership of $i$ in $\mathcal{B}$ is monotone in $v_i$, and that the truthful payment for $i \in \mathcal{B}$ is precisely $B_i$, since this is the minimum value required for allocation. Thus, we can immediately see that

$$\mathcal{R}^{M_\mathcal{B}} = \mathop{\mathrm{E}}_{\hat{\mathbf{v}} \sim \widehat{\mathbf{F}}} \left[ \sum_{i \in \mathcal{B}} B_i \right] = \mathop{\mathrm{E}}_{\mathbf{v} \sim \mathbf{F}} \left[ \sum_{i \in \mathcal{B}} B_i \right],$$

as desired. □

By combining the results of Lemmas 5 and 6, we get the following theorem.

**Theorem 7.** *Given a single parameter setting $\mathcal{I} = (\mathbf{F}, \mathcal{S}, \mathbf{B})$, the optimal mechanism $\mathcal{M}$ for the modified setting $\widehat{\mathcal{I}} = (\widehat{\mathbf{F}}, \mathcal{S})$ gives a 2-approximation to the optimal revenue for $\mathcal{I}$.*

### 3.3 Multi-parameter setting with public budgets

We next consider settings where a seller offers multiple kinds of service and agents have different preferences over them. Agents are unit-demand and want any one of the services; the seller faces a general downward closed feasibility constraint. As before, we use the tuple $\mathcal{I} = (\mathbf{F}, \mathcal{S}, \mathbf{B})$ to denote an instance of this problem; throughout, $i$ indexes agents and $j$ indexes services. Let $\mathcal{S}$ be a downward-closed feasibility constraint over $(i, j)$ pairs, and furthermore assume each agent $i$ has a budget $B_i$.

Ideally, we would like to follow the same approach as in the previous section. We use the same basic benchmark, defining $\widehat{\mathbf{F}}$ and $\mathcal{B}$ analogously to (1) and (2) for the instance $\mathcal{I}$. Note that here, $\mathcal{B}$ is a collection of $(i, j)$ pairs; since agents are unit-demand, however, we can also think of $\mathcal{B}$ as a set of agents. We can't apply the same reduction from $M$ to $\widehat{M}$ directly, however, because truncating each of a multi-parameter agent's values to their budget affects the agents' preferences across different items, a concern we did not have before.

Instead, we make use of a reduction of Chawla et al. [6] from multi-parameter Bayesian MD to single-parameter Bayesian MD to first bring the problem into a single parameter domain and then apply the approach from the previous section. We begin by briefly describing the reduction of Chawla et al. [6]; the reader is referred to [6] for details.

- For an instance $\mathcal{I}$, [6] defines a single-parameter instance $\mathcal{I}^{\text{reps}}$, where each original agent is split into multiple "representatives", each representing the agent's interest in exactly one of the available services. In the setting $\mathcal{I}^{\text{reps}}$, the single-parameter representatives compete against each other.

- [6] show that any deterministic mechanism $M$ for $\mathcal{I}$ can be converted into one, say $M'$, for $\mathcal{I}^{\text{reps}}$ that has the same allocation rule as $M$ but induces larger truthful payments.

- [6] show that for certain feasibility constraints (notably, matroids and matroid intersections), there exist approximately optimal mechanisms for $\mathcal{I}^{\text{reps}}$ that can be converted into mechanisms for $\mathcal{I}$ without loss in revenue or truthfulness.

We now show how to carry out the reduction of Chawla et al. in the context of budgets. Specifically, starting with a budget feasible mechanism $M$ for $\mathcal{I}$, we first follow the approach in [6] to convert it into a mechanism $M'$ for $\mathcal{I}^{\text{reps}}$ with the same allocation rule as for $M$. Unfortunately, $M'$ is not necessarily budget feasible. We therefore modify the mechanism so that any representative in $\mathcal{I}^{\text{reps}}$ offered service at a price larger than his budget is dropped from the allocated set. This makes $M'$ budget feasible. We then apply the approach of the previous section to construct a mechanism $\widehat{M}$ for the instance $\widehat{\mathcal{I}}^{\text{reps}}$ based on the modified $M'$.

Finally, we apply Chawla et al.'s approach to the setting $\widehat{\mathcal{I}}^{\text{reps}}$ to obtain an approximately-optimal *sequential posted price* mechanism; this mechanism is necessarily budget-feasible because in the setting $\widehat{\mathcal{I}}^{\text{reps}}$ values do not exceed budgets. We then apply this mechanism (with the same allocation and payment rules) to the original setting $\mathcal{I}$. We call this final mechanism **S**.

**Lemma 8.** *Consider a multi-parameter setting $\mathcal{I} = (\mathbf{F}, \mathcal{S}, \mathbf{B})$, and let $\alpha$ denote the approximation to revenue that Chawla et al.'s mechanism achieves in this setting. Then for any deterministic budget feasible mechanism $M$, the mechanism **S** defined above satisfies*

$$\mathcal{R}^{\mathbf{S}} \geq 1/\alpha \operatorname*{E}_{\mathbf{v}} \left[ \sum_{i \in L} \mathcal{R}_i^M(\mathbf{v}) \right]$$

*and consequently,*

$$\mathcal{R}^M \leq \alpha \mathcal{R}^{\mathbf{S}} + \operatorname*{E}_{\mathbf{v} \sim \mathbf{F}} \left[ \sum_{i \in \mathcal{B}} B_i \right].$$

*Proof.* We first note that the revenue that $M'$ derives from the agents in $L$ is no smaller than the revenue that $M$ derives from these agents because payments for these agents never exceed their budgets. Following the analysis of Lemma 5, we further conclude that the revenue of $\widehat{M}$ is an upper bound on the contribution of agents in $L$ to $\mathcal{R}^M$. Finally, the statement of the lemma implies that the revenue of the mechanism **S** is at least as large as a $1/\alpha$ fraction of $\mathcal{R}^{\widehat{M}}$. This proves the first part of the lemma. The second statement follows along the lines of the proof of Lemma 5. □

Finally, we show how to approximate the benchmark $\mathrm{E}[\sum_{i \in \mathcal{B}} B_i]$ in the multi-parameter setting.

**Lemma 9.** *If $\mathcal{S}$ is a matroid set system, there exists a mechanism $M_{\mathcal{B}}$ for the setting $\widehat{\mathcal{I}}$ such that $\mathrm{E}_{\mathbf{v} \sim \mathbf{F}} \left[ \sum_{i \in \mathcal{B}} B_i \right] \leq 2\mathcal{R}^{M_{\mathcal{B}}}$.*

*Proof.* We begin by noting that the unit-demand constraint on agents is precisely a partition matroid; furthermore, taking a subset of a matroid induces a matroid. Hence for any fixed $\mathbf{v}$, our objective $\sum_{i \in \mathcal{B}} B_i$ is precisely a maximum weighted set in the intersection of two matroids. Note that each $(i, j)$ that is a valid element of $\mathcal{B}$ corresponds to an agent $i$ who would be willing to pay a price of $B_i$ for service $j$.

Our proposed mechanism $M_{\mathcal{B}}$ sequentially approaches each agent in order by decreasing $B_i$, and offers them all services that are still feasible under $\mathcal{S}$ (based on previous decisions by agents), at a price of $B_i$. Then our revenue is the weight of a greedily selected independent set in the matroid intersection $\mathcal{B}$ is optimal over; by Theorems 1.1 and 3.2 of [13], this set's weight is at least $1/2$ that of $\mathcal{B}$. Note that as a sequential posted pricing, the mechanism is immediately truthful; taking expectations over $\mathbf{v}$ yields the claimed revenue bound. □

Combining Lemmas 8 and 9 immediately gives us the following theorem.

**Theorem 10.** *Let $\mathcal{I} = (\mathbf{F}, \mathcal{S}, \mathbf{B})$ be an instance with multi-parameter, unit-demand agents and $\mathcal{S}$ being a matroid or simpler feasibility constraint. Then, there exists a polynomial time computable mechanism for $\widehat{\mathcal{I}}$ that is budget feasible and DSIC for $\mathcal{I}$ and obtains a constant fraction of the revenue of the optimal budget-feasible mechanism for $\mathcal{I}$.*

### 3.4 Private budgets

We next consider settings where budgets are part of agents' private types, but where the mechanism designer knows the distributions from which budgets are drawn. We assume that values and budgets are drawn from independent distributions. We focus on settings where agents' values are single-dimensional.

Let $\mathcal{I} = (\mathbf{F}, \mathcal{S}, \mathbf{G})$ denote an instance of this setting. We follow a similar approach as for public budgets. Switching from public to private budgets, however, adds new complexity; in particular it becomes tricky to achieve our benchmark $\mathrm{E}[\sum_{i \in \mathcal{B}} B_i]$. In this section, we present an approximately optimal mechanism for the case when each distribution in $\mathbf{F}$ satisfies the MHR condition (see Definition 1 in Section 2). Our analysis uses the MHR condition in a very mild way and in fact holds for any setting where for every agent the probability that his value exceeds his monopoly price is lower bounded by a constant. Even when this condition is not satisfied, we can obtain a good approximation through a slight modification of our mechanism under a technical condition on values and budgets. We believe that the general idea behind our mechanism can be extended to obtain good approximations for arbitrary distributions.

We focus on settings $\mathcal{I} = (\mathbf{F}, \mathcal{S}, \mathbf{G})$ where $\mathcal{S}$ is a matroid set system, and each distribution in $\mathbf{F}$ satisfies the MHR condition. We begin with some definitions. Given a pair of value and budget vectors, we consider the "extractable value" of an agent $i$ to be $\min(v_i, B_i)$; we modify our definition of $\mathcal{B}$ to reflect this:

$$\mathcal{B} = \mathrm{argmax}_{S \in \mathcal{S}} \sum_{i \in S} \min(v_i, B_i).$$

Similarly to the public budgets case, our approach is to split the revenue of an arbitrary mechanism into two terms, which (loosely speaking) look like revenue in a truncated value setting, and the sum of the budgets in $\mathcal{B}$; we then demonstrate a lottery menu mechanism whose revenue upper bounds both of these terms.

Our proposed mechanism (which we denote $M^{\mathcal{L}}$) is based on lottery menus $\mathcal{L}(\underline{p}, \bar{p})$ parameterized by a minimum price $\underline{p}$ and a maximum price $\bar{p}$. (Recall the discussion of lotteries at the beginning of this section.) There are two cases. If $\underline{p} \geq \bar{p}/3$, then $\mathcal{L}(\underline{p}, \bar{p})$ contains the single fixed price of $\underline{p}$; otherwise, it contains, for all $2\underline{p}/\bar{p} \leq \alpha \leq 2/3$, a lottery that with probability $(1/3 + \alpha)$ allows the agent to purchase service at a price of $\alpha \bar{p}/2$.

Note that the probability of allocation in the above lottery system rises faster as a function of $\alpha$ than the price of the lottery. Effectively this ensures that the agent is willing to buy the most expensive lottery that he can afford. So, in particular, if all lotteries bring positive utility then the agent spends his entire budget, the maximum amount that any mechanism can hope to achieve from the agent. This powerful idea is what enables our approximation.

Note the following properties of the lottery menu $\mathcal{L}(\underline{p}, \bar{p})$.

**Lemma 11.** *When an agent with $\min(v, B) \geq \underline{p}$ is offered the menu $\mathcal{L}(\underline{p}, \bar{p})$, he purchases an option yielding expected revenue at least $\underline{p}/3$.*

*Proof.* Note that in either case, the lottery system $\mathcal{L}(\underline{p}, \bar{p})$ always contains an option with price exactly $\underline{p}$, and that this is the lowest priced option. Furthermore, for $\min(v, B) \geq \underline{p}$, this option always gives non-negative utility. Thus our claim follows immediately from the fact that every lottery assigns the item with probability at least $1/3$. □

**Lemma 12.** *When an agent with $v \geq \bar{p}$ is offered the menu $\mathcal{L}(\underline{p}, \bar{p})$, he purchases an option yielding expected revenue at least $\min(\bar{p}, B)/3$.*

*Proof.* In the first case (when $\mathcal{L}(\underline{p}, \bar{p})$ contains the single price of $\underline{p}$), this follows trivially; in the other case, consider the utility of an agent with value $v$ when purchasing the lottery with parameter $\alpha$, which we denote $u_\alpha(v)$. We can see that

$$u_\alpha(v) = (1/3 + \alpha)(v - \alpha \bar{p}/2), \text{ and so}$$
$$\frac{\partial u_\alpha(v)}{\partial \alpha} = v - (\alpha + 1/6)\bar{p} \geq 0$$

by our assumption. So we can see that an agent will purchase the lottery with the highest $\alpha$ value they can afford; since the lottery for $\alpha = 2/3$ assigns service with probability 1 at a price of $\bar{p}/3$, and every lottery provides service with probability at least one third, we can see that an agent will purchase a lottery yielding revenue at least $\min(\bar{p}, B)/3$. □

Our mechanism $M^{\mathcal{L}}$ serves the set $\mathcal{B}$. For each $i$, let $T_i$ be the threshold corresponding to inclusion in $\mathcal{B}$, i.e. $T_i = \min\{v' : i \in \mathcal{B} \text{ for } ((\mathbf{v}_{-i}, v'), (\mathbf{B}_{-i}, v'))\}$. Our mechanism offers agent $i$ the lottery system $\mathcal{L}(T_i, \phi_i^{-1}(0))$, where $\phi_i^{-1}(0)$ is the monopoly price for $i$.

In order to relate the revenue of a mechanism $M$ for the setting $\mathcal{I}$ to that of our proposed mechanism $M^{\mathcal{L}}$, we break the revenue of $M$ into two parts – that derived from agents in $\mathcal{B}$, and that derived from agents not in $\mathcal{B}$; we denote these quantities by $\mathcal{R}^{M \cap \mathcal{B}}$ and $\mathcal{R}^{M \setminus \mathcal{B}}$, respectively. We bound the two terms separately.

**Lemma 13.** $\mathcal{R}^{M \setminus \mathcal{B}} \leq 3\mathcal{R}^{M^{\mathcal{L}}}$

*Proof.* Let $S$ denote the set of agents served by $M$. Note that by our definition of $\mathcal{B}$, it will always be a maximal independent set in $\mathcal{S}$; as such, for every pair of vectors $(\mathbf{v}, \mathbf{B})$, we can get a 1-1 function $g : S \setminus \mathcal{B} \to \mathcal{B}$ such that for all $i \in S \setminus \mathcal{B}, \mathcal{B} \setminus \{g(i)\} \cup \{i\} \in \mathcal{S}$. Note that by EPIR, we must have that the revenue $M$ derives from each agent $i$ is no more than $\min(v_i, B_i)$, and by the definition of $T_i$, we get that

$$\mathcal{R}^{M \setminus \mathcal{B}}(\mathbf{v}, \mathbf{B}) \leq \sum_{i \in S \setminus \mathcal{B}} \min(v_i, B_i) \leq \sum_{i \in S \setminus \mathcal{B}} T_{g(i)} \leq \sum_{i \in \mathcal{B}} T_i \leq 3 \mathcal{R}^{M^{\mathcal{L}}}(\mathbf{v}, \mathbf{B})$$

Here the second inequality follows from noting that in order to be in the set $\mathcal{B}$, the agent $g(i)$ must have an extractable value no smaller than that of $i$. The last inequality follows from applying Lemma 11 to $\mathcal{L}(T_i, \phi_i^{-1}(0))$. Taking expectation over $(\mathbf{v}, \mathbf{B})$ completes the proof. □

In order to prove our next revenue bound, we need the following property of MHR variables (Lemma 4.1 in [11]):

**Lemma 14.** *For $v$ distributed according to some $F$ satisfying the MHR, the probability that the value $v$ exceeds the monopoly price $\phi^{-1}(0)$ is at least $1/e$.*

**Lemma 15.** $\mathcal{R}^{M \cap \mathcal{B}} \leq 3e \mathcal{R}^{M^{\mathcal{L}}}$

*Proof.* Consider some agent $i$, and fix $(\mathbf{v}_{-i}, \mathbf{B}_{-i})$; note that this fixes $T_i$ as well. Fix $B_i$. Recall that an agent $i$ contributes to $\mathcal{R}_i^{M \cap \mathcal{B}}$ only when $\min(v_i, B_i) \geq T_i$. We split the analysis into two cases; in each case we consider the optimal revenue a mechanism could derive from agents with $v_i \geq T_i$ if allowed to ignore the budgets constraints.

- Case 1: $T_i \geq \phi_i^{-1}(0)$. In this case, the maximum revenue that can be obtained from agent $i$ conditioned on $v_i \geq T_i$ and ignoring feasibility constraints is $T_i$ and can be obtained by offering a fixed price of $T_i$. Our mechanism on the other hand offers a single option of buying service at a fixed price of $T_i$. Therefore,

$$\mathop{\mathrm{E}}_{v_i}\left[\mathcal{R}_i^{M \cap \mathcal{B}}(\mathbf{v}, \mathbf{B}) \,|\, v_i \geq T_i\right] \leq \mathop{\mathrm{E}}_{v_i}\left[\mathcal{R}^{M^{\mathcal{L}}}(\mathbf{v}, \mathbf{B}) \,|\, v_i \geq T_i\right]$$

- Case 2: $T_i < \phi_i^{-1}(0)$. The maximum revenue that can be obtained from agent $i$ conditioned on $v_i \geq T_i$ and ignoring feasibility constraints is at most $\phi_i^{-1}(0)$ and can be obtained by offering a fixed price of $\phi_i^{-1}(0)$. Considering the budget constraint we conclude that $\mathrm{E}_{v_i}[\mathcal{R}_i^{M \cap \mathcal{B}}(\mathbf{v}, \mathbf{B}) \,|\, v_i \geq T_i] \leq \min(\phi_i^{-1}(0), B_i)$. On the other hand, applying Lemma 12 to $\mathcal{L}(T_i, \phi_i^{-1}(0))$, we get that for $v_i \geq \phi_i^{-1}(0)$, $\mathcal{R}_i^{M^{\mathcal{L}}} \geq \min(\phi_i^{-1}(0), B_i)/3$. Lemma 14 implies that the event $v_i \geq \phi_i^{-1}(0)$ happens with probability at least $1/e$. So we get

$$\mathop{\mathrm{E}}_{v_i}[\mathcal{R}_i^{M \cap \mathcal{B}}(\mathbf{v}, \mathbf{B})\,|\, v_i \geq T_i] \leq (1/3e) \mathop{\mathrm{E}}_{v_i}[\mathcal{R}^{M^{\mathcal{L}}}(\mathbf{v}, \mathbf{B}) \,|\, v_i \geq T_i]$$

We have $\mathrm{E}_{v_i}[\mathcal{R}_i^{M \cap \mathcal{B}}(\mathbf{v}, \mathbf{B})] \leq 3e \, \mathrm{E}_{v_i}[\mathcal{R}^{M^{\mathcal{L}}}(\mathbf{v}, \mathbf{B})]$ in either case; taking expectations over $(\mathbf{v}_{-i}, \mathbf{B})$, and summing over $i$ yields our claim. □

Combining the above two lemmas immediately gives the following theorem.

**Theorem 16.** *For any setting $\mathcal{I} = (\mathbf{F}, \mathcal{S}, \mathbf{G})$ where $\mathcal{S}$ is a matroid constraint and each distribution in $\mathbf{F}$ satisfies MHR, the mechanism $M^{\mathcal{L}}$ is a $3(1 + e)$ approximation to the optimal revenue.*

In fact, we can use a similar lottery pricing technique to get a constant approximation even in the absence of the MHR assumption; however, we still need a technical assumption relating the distributions of agents' values and budgets. This once again ensures that there is a good probability that agents' values are high enough for the lottery system to extract a constant fraction of their budget. We state the theorem here, but defer the proof of this result to Appendix C.

**Theorem 17.** *Suppose that every agent's median value is no smaller than a constant fraction of her maximum budget. Then we can construct a budget-feasible mechanism that is DSIC with respect to both values and budgets, and obtains a constant fraction of the revenue of the optimal such mechanism.*

# 4 Maximizing welfare

In this section we focus on the welfare objective. In particular, the seller's goal is to maximize the total value of the allocation in expectation. Once again we assume that budgets are known publicly.

We first note that we cannot use the approach of the previous section as a roadmap. Even with public budgets, truncating values to the corresponding budgets does not work for the social welfare objective. In particular, the following example shows it is possible for a budget feasible mechanism to distinguish between values above the budget without exceeding the budget in payments.

**Example 2.** *Consider an $n$ agent single-item auction, where agents have i. i. d. values for the item. Each agent has a budget of $1$. Each agent's value is $1$ with probability $1 - 1/n$ and $n$ with probability $1/n$. Then a mechanism that simply truncates values to budgets cannot distinguish between the agents and gets a social welfare of at most $2$. On the other hand, consider a mechanism that orders agents in an arbitrary order and offers two options to each agent in turn while the item is unallocated: getting the item for free with probability $1/n$ and nothing otherwise, or purchasing the item at a price of $1$. Then, an agent picks the first option if and only if her value is below $n/(n-1)$, and otherwise picks the second option. In particular, an agent with value $n$ always picks the second option, and an agent with value $1$ always picks the first option. For large $n$, with probability approaching $1 - 1/e$ at least one agent has value $n$, and with probability at least $1/e$ the item is unsold before the first agent with value $n$ is made an offer. The mechanism's expected welfare is therefore at least $1/e(1 - 1/e)n = \Omega(n)$.*

Note that the precise choice of budgets in the above example was critical: if budgets were any lower, the proposed mechanism would have been infeasible; and if they were any higher, truncation would have still allowed for distinguishing between agents with low and high values. This suggests considering bicriteria approximations where we compete against an optimal mechanism that faces smaller budgets. We first demonstrate a mechanism achieving an approximation of this sort; we then show that our mechanism also gives a good approximation if we relax the EPIR constraint to an IIR constraint, instead of relaxing budgets. Of course, our ultimate goal is to provide a good approximation for the social welfare objective via an EPIR budget feasible mechanism. While we are unable to do so in general, the final section presents a constant factor approximation for settings where the distributions $F_i$ for every agent $i$ satisfy the MHR condition (Definition 1 in Section 2).

## 4.1 A bicriteria approximation

Consider a setting $\mathcal{I}$ with budgets $\mathbf{B}$. Let OPT$'$ denote the EPIR mechanism that is welfare-maximizing and feasible for budgets $(1-\epsilon)\mathbf{B}$ (i.e. where each budget is scaled down by a factor of $1-\epsilon$). We claim that we can approximate the welfare of this mechanism while maintaining budget feasibility with respect to the original budgets $\mathbf{B}$.

**Theorem 18.** *For a given instance $\mathcal{I} = (\mathbf{F}, \mathcal{S}, \mathbf{B})$, let $\mathcal{I}'$ be the instance $(\mathbf{F}, \mathcal{S}, (1-\epsilon)\mathbf{B})$ where each agent's budget is scaled down by a factor of $1 - \epsilon$. Let OPT$'$ denote the welfare optimal budget feasible mechanism for $\mathcal{I}'$. Then, there exists an easy to compute ex post IR mechanism (namely, the VCG mechanism over a modified instance) that is budget feasible for $\mathcal{I}$ and obtains at least an $\epsilon$ fraction of the social welfare of OPT$'$.*

*Proof.* We first use OPT$'$ to construct a new mechanism $M$. $M$ proceeds as follows. It elicits values from agents. For all agents $i$ with $v_i \geq B_i$, it resamples agent $i$'s value from the distribution $F_i$ restricted to the set $[B_i, \infty)$. Other values are left unchanged. It then runs the mechanism OPT$'$ on the resampled values. It is easy to see that $M$ is budget feasible valid for $\mathcal{I}'$.

We claim that the social welfare of $M$ is at least $\epsilon$ times the social welfare of OPT$'$. To prove the claim, consider a single agent $i$, and let $x_i^1$ denote the probability of allocation for this agent in OPT$'$ when her value is $B_i$, and $x_i^2$ denote the probability of allocation for this agent in OPT$'$ when her value is $v_i^{\max}$ (the agent's maximum possible value). Note that the expected payment that the agent makes at $v_i^{\max}$ is at least $(x_i^2 - x_i^1)B_i$ plus the payment she makes at $B_i$. Then, EPIR implies that $(x_i^2 - x_i^1)B_i$ is at most the budget $(1-\epsilon)B_i$ times $x_i^2$. This implies $x_i^2 < x_i^1/\epsilon$. Now, noting that the value distributions for agents are unaltered by resampling, it holds that for $v_i \geq B_i$, the probability of allocation for agent $i$ at $v_i$ under $M$ is equal to the expected probability of allocation for the agent under OPT$'$ conditioned on the agent's value being in the range $[B_i, \infty)$. Since the probability of allocation under OPT′ for this range is always between $x_i^1$ and $x_i^2$, the expected probability of allocation is at least $x_i^1 \geq \epsilon x_i^2$. So compared to those under $OPT'$, the probabilities of allocation under $M$ are at most a factor of $\epsilon$ smaller. Therefore, the expected social welfare of $M$ is also at most a factor of $\epsilon$ smaller than that of OPT′.

Our goal will then be to approximate the social welfare of $M$. Note that for every agent $i$, $M$ treats values above $B_i$ identically. We can therefore consider the following optimization problem: for an instance $\mathcal{I}$, construct a DSIC EPIR mechanism that maximizes social welfare subject to the additional constraint that for every agent $i$ the mechanism's (distribution over) allocation should be identical across value vectors that differ only in agent $i$'s value and where agent $i$'s value is $\geq B_i$. For any such mechanism, agent $i$'s expected contribution to social welfare from value vectors with $v_i \geq B_i$ conditioned on being allocated is $\overline{v}_i$, where $\overline{v}_i = E[v_i|v_i \geq B_i]$. Therefore, the following mechanism maximizes welfare over the above class of mechanisms: for every agent $i$ with $v_i \geq B_i$, replace $v_i$ by $\overline{v}_i$; other values remain unmodified; run the VCG mechanism over the modified value vector; charge every agent the minimum of the payment returned by the VCG mechanism and their budget. It is easy to verify that this mechanism is DSIC, ex post IR, budget feasible for the original budgets $B_i$, and obtains expected social welfare at least that of $M$. Therefore, it satisfies the claim in the theorem. □

## 4.2 An interim IR mechanism

Next we note that it is in fact easy to remove the approximation on budget in the above theorem if we are willing to give up on EPIR. In particular, consider an optimal mechanism OPT on the instance $\mathcal{I} = (\mathbf{F}, \mathcal{S}, \mathbf{B})$. Then the above theorem implies the existence of a mechanism $V$ that is budget feasible for $\mathcal{I}' = (\mathbf{F}, \mathcal{S}, 2\mathbf{B})$ and obtains half the welfare of OPT (taking $\epsilon = 1/2$). Now consider the mechanism $V'$ described as follows. $V'$ simulates $V$ on the given value vector. Then for every agent $i$ it charges $i$ half the payment charged by $V$ and with probability $1/2$ makes an allocation to $i$ if $V$ makes an allocation to $i$. Agent $i$'s expected utility from any strategy under $V'$ is exactly half her expected utility from the same strategy under $V$. Therefore, $V'$ is DSIC. Moreover, it is budget feasible for the original budgets $\mathbf{B}$ since it always charges half the payments in $V$. Its expected social welfare is exactly half that of $V$. We therefore get the following theorem.

**Theorem 19.** *For a given instance $\mathcal{I} = (\mathbf{F}, \mathcal{S}, \mathbf{B})$, let OPT denote the welfare optimal EPIR budget feasible mechanism for $\mathcal{I}$. Then, there exists an easy to compute IIR mechanism that is budget feasible for $\mathcal{I}$ and obtains at least a quarter of the social welfare of OPT.*

## 4.3 An ex-post IR mechanism for MHR distributions

As previously remarked, our ultimate goal is to provide a good approximation for the social welfare objective via an EPIR budget feasible mechanism. We now show that under an MHR condition on distributions, we can achieve precisely this goal. In particular, we present a constant factor approximation for settings where the distributions $F_i$ for every agent $i$ satisfy the MHR condition (Definition 1 in Section 2).

Under the MHR condition, we can exhibit a close relationship between the welfare and revenue of any mechanism. Using this relationship along with results from the previous section, we can come up with a budget feasible approximately-revenue-maximizing mechanism that also provides an approximation to social welfare. The MHR condition is quite crucial to our approach. In fact our solution consists of two mechanisms, one of which charges no payments, and the other of which truncates values to their corresponding budgets – approaches that don't work for the example we considered above.

Let $v_i^* = \phi_i^{-1}(0)$ denote the monopoly price for the distribution $F_i$. We then get the following bound on social welfare, which we are able to approximate.

**Lemma 20.** *For any instance $\mathcal{I} = (\mathbf{F}, \mathcal{S}, \mathbf{B})$, if all the distributions $F_i$ satisfy the MHR condition, then for any non-decreasing allocation function $x(v)$, we have that*

$$\int_{\mathbf{v}} \left(\sum_i v_i x_i(\mathbf{v})\right) d\mathbf{F}(\mathbf{v}) \leq \int_{\mathbf{v}} \left(\sum_i (\phi_i(v_i) + 2v_i^*) x_i(\mathbf{v})\right) d\mathbf{F}(\mathbf{v}).$$

In order to prove the Lemma, we require some new definitions and claims. Consider a single agent with MHR distribution $F$, virtual value function $\phi$, and monopoly price $v^*$. Let $\phi^+$ and $\phi^-$ be the positive and negative portions of $\phi$ respectively; i.e. for all $v$, $\phi^+(v), \phi^-(v) \geq 0$ and $\phi(v) = \phi^+(v) - \phi^-(v)$. We can then claim the following (the first is a restatement of Lemma 3.1 in [12]).

**Lemma 21.** *For a distribution $F$ satisfying the MHR, all values $v$ satisfy $v \leq v^* + \phi^+(v) = v^* + \phi(v) + \phi^-(v)$.*

**Lemma 22.** *For any monotone allocation function $x(\cdot)$,*

$$\int \phi^-(v) x(v) dF(v) \leq \int v^* x(v) dF(v).$$

*Proof.* We begin by recalling that by Lemma 1, the expected revenue of any BIC mechanism is equal to its expected virtual surplus. Now consider a mechanism for a single agent with value distribution $F$ that always serves the agent. Clearly the revenue of this mechanism is 0. So we get

$$\int_v (\phi^+(v) - \phi^-(v)) dF(v) = 0,$$

which implies that

$$\int_v \phi^+(v) dF(v) = \int_v \phi^-(v) dF(v).$$

Second, the revenue from offering the agent the monopoly price $v^*$ is precisely $v^*(1 - F(v^*))$. Therefore,

$$\int \phi^+(v) dF(v) = \int_{v^*}^{\infty} (\phi^+(v) - \phi^-(v)) dF(v) = v^* \int_{v^*}^{\infty} dF(v),$$

where the first equality follows from regularity of $F$.

Note that the regularity of $F$ implies that $\phi^+$ and $\phi^-$ are identically 0 below and above $v^*$, respectively. Now, from the above two equalities, we can see that if $x$ is monotone non-decreasing, then,

$$\int \phi^-(v) x(v) dF(v) \leq \int \phi^-(v) x(v^*) dF(v) = x(v^*) \int \phi^+(v) dF(v) = x(v^*) v^* \int_{v^*}^{\infty} dF(v) \leq \int v^* x(v) dF(v);$$

the claim follows. □

The proof of Lemma 20 follows immediately by combining Lemmas 21 and 22. The Lemma gives us the following approximation.

**Theorem 23.** *Let $\mathcal{I} = (\mathbf{F}, \mathcal{S}, \mathbf{B})$ be an instance where all distributions $F_i$ satisfy the MHR condition. Then, one of the following mechanisms obtains a $2(1+e)$-approximation to the social welfare of a welfare-optimal budget-feasible mechanism for $\mathcal{I}$. Both of these mechanisms are DSIC, EPIR and budget feasible.*

- *Mechanism 1: Always allocate to the set $S_1^*$ and charge zero payments, where $S_1^* = \mathrm{argmax}_{S \in \mathcal{S}} \sum_{i \in S} v_i^*$.*

- *Mechanism 2: Elicit values from agents; for all $i$ with $v_i > B_i$, replace $v_i$ by $B_i$; run Myerson's mechanism on the resulting instance.*

*Proof.* We begin by noting that an immediate consequence of Lemma 14 is that for a distribution $F$ satisfying the MHR, $E_F[v] \geq v^*/e$.

Now, consider some budget feasible mechanism $M$ for the instance $\mathcal{I}$. Then by Theorem 7, the optimal mechanism for the truncated distributions (1) obtains revenue, and therefore also social welfare, no less than a $1/2$ fraction of the expected revenue $\int \sum_i \phi_i(v_i) x_i(\mathbf{v}) d\mathbf{F}(\mathbf{v})$.

Moreover, $\int_{\mathbf{v}} \sum_i v_i^* x(\mathbf{v}) d\mathbf{F}(\mathbf{v})$ is upper bounded by $\sum_{i \in S^*} v_i^*$, where $S^* = \mathrm{argmax}_{S \in \mathcal{S}} \sum_{i \in S} v_i^*$. Then a mechanism which always allocates to the set $S^*$ and charges no payments is budget feasible, DSIC, and obtains welfare $\sum_{i \in S^*} E_{F_i}[v_i] \geq 1/e \sum_{i \in S^*} v_i^*$, where the inequality follows from Lemma 14. The original claim follows. □

## Acknowledgments

We thank Saeed Alaei for many helpful discussions.

## References

[1] Z. Abrams. Revenue maximization when bidders have budgets. In *Proceedings of the seventeenth annual ACM-SIAM Symposium on Discrete Algorithms*, SODA '06, pages 1074–1082, New York, NY, USA, 2006. ACM.

[2] S. Alaei, K. Jain, and A. Malekian. Competitive equilibrium for unit-demand buyers with non quasi-linear utilities. In *CORR*, 2010.

[3] S. Bhattacharya, G. Goel, S. Gollapudi, and K. Munagala. Budget constrained auctions with heterogeneous items. In *STOC*, pages 379–388, 2010.

[4] C. Borgs, J. T. Chayes, N. Immorlica, M. Mahdian, and A. Saberi. Multi-unit auctions with budget-constrained bidders. In *ACM Conference on Electronic Commerce*, pages 44–51, 2005.

[5] S. Brusco and G. Lopomo. Simultaneous ascending auctions with complementarities and known budget constraints. *Economic Theory*, 38(1):105–124, January 2009.

[6] S. Chawla, J. D. Hartline, D. L. Malec, and B. Sivan. Multi-parameter mechanism design and sequential posted pricing. In *STOC*, pages 311–320, 2010.

[7] Y.-K. Che and I. Gale. Standard auctions with financially constrained bidders. *Review of Economic Studies*, 65(1):1–21, January 1998.

[8] Y.-K. Che and I. Gale. The optimal mechanism for selling to a budget-constrained buyer. *Journal of Economic Theory*, 92(2):198–233, June 2000.

[9] N. Chen, X. Deng, and A. Ghosh. Competitive equilibria in matching markets with budgets. *CoRR*, 2010.

[10] S. Dobzinski, N. Nisan, and R. Lavi. Multi-unit auctions with budget limits. In *Proc. of the 49th Annual Symposium on Foundations of Computer Science (FOCS)*, 2008.

[11] J. Hartline, V. Mirrokni, and M. Sundararajan. Optimal marketing strategies over social networks. In *Proceeding of the 17th international conference on World Wide Web*, WWW '08, pages 189–198, New York, NY, USA, 2008. ACM.

[12] J. Hartline and T. Roughgarden. Simple versus optimal mechanisms. In *Proc. 11th ACM Conf. on Electronic Commerce*, 2009.

[13] B. Korte and D. Hausmann. An analysis of the greedy heuristic for independence systems. In P. H. B. Alspach and D. Miller, editors, *Algorithmic Aspects of Combinatorics*, volume 2 of *Annals of Discrete Mathematics*, pages 65 – 74. Elsevier, 1978.

[14] J.-J. Laffont and J. Robert. Optimal auction with financially constrained buyers. *Economics Letters*, 52(2):181–186, 1996.

[15] E. S. Maskin. Auctions, development, and privatization: Efficient auctions with liquidity-constrained buyers. *European Economic Review*, 44(4-6):667–681, 2000.

[16] R. Myerson. Optimal auction design. *Mathematics of Operations Research*, 6:58–73, 1981.

[17] N. Nisan. Introduction to mechanism design. In N. Nisan, T. Roughgarden, E. Tardos, and V. V. Vazirani, editors, *Algorithmic Game Theory*, chapter 9, pages 209–242. Cambridge Press, 2007.

[18] N. Nisan, J. Bayer, D. Chandra, T. Franji, R. Gardner, Y. Matias, N. Rhodes, M. Seltzer, D. Tom, H. Varian, and D. Zigmond. Google's auction for tv ads. In *Proceedings of the 36th International Colloquium on Automata, Languages and Programming: Part II*, ICALP '09, pages 309–327, Berlin, Heidelberg, 2009. Springer-Verlag.

[19] M. M. Pai and R. Vohra. Optimal auctions with financially constrained bidders. Discussion papers, Northwestern University, Aug. 2008.

## A  Incentive compatibility for budgets

Let $\mathcal{I} = (\mathbf{F}, \mathcal{S}, \mathbf{G})$ be a mechanism design instance with single-parameter agents with private budgets, and let $M$ be a truthful mechanism for this setting. We get the following characterization for the allocation and payment functions of $M$.

**Lemma 24.** *If $M$ is truthful, then for each $i$ there exists some monotone function $\tilde{x}_i(v)$ such that for each $B$, $x_i(v, B)$ has the form*

$$x_i(v, B) = \begin{cases} \tilde{x}_i(v) & \text{if } v < v_B \\ \tilde{x}_i(v_B) & \text{if } v \geq v_B, \end{cases}$$

*where $v_B$ is a monotone non-decreasing function of $B$, and the payment has the form*

$$p_i(v, B) = x_i(v, B)v - \int_0^v x_i(t, B) dt.$$

*Proof.* For now, assume that $x_i(v, B)$ and $p_i(v, B)$ are continuous. Begin by fixing some arbitrary budget $B$, and considering the truthfulness constraints on $v$. Just as in the case where we have no budgets, truthfulness in reporting valuation implies that $x_i(v, B)$ is monotone increasing, and payments are of the form

$$p_i(v, B) = x_i(v, B)v - \int_0^v x_i(t, B) dt;$$

so we need only show the claimed relation between allocation curves for different budgets. Denote the utility $i$ receives by $u_i(v, B)$. Note by the above, we get that

$$\begin{aligned} u_i(v, B) &= x_i(v, B)v - p_i(v, B) \\ &= x_i(v, B)v - \left( x_i(v, B)v - \int_0^v x_i(t, B) dt \right) \\ &= \int_0^v x_i(t, B) dt. \end{aligned}$$

Begin by fixing some $v$, and considering two different budgets $B$ and $B'$ (without loss of generality $B < B'$). Since an agent with budget $B'$ must be able to afford any payment that an agent with budget $B$ can, we know that $u_i(v, B') \geq u_i(v, B)$. Assume that $u_i(v, B') = u_i(v, B)$; then we have that $x_i(v, B') = x_i(v, B)$ as well. Suppose not; without loss of generality, say $x_i(v, B') > x_i(v, B)$. Then by continuity, this holds on some neighborhood $(v - \delta, v + \delta)$ of $v$. But then we get that

$$\begin{aligned} u_i(v - \delta, B') &= \int_0^{v-\delta} x_i(t, B') dt \\ &= \int_0^{v-\delta} x_i(t, B) dt + \int_{v-\delta}^v x_i(t, B) - x_i(t, B') dt \\ &< u_i(v - \delta, B), \end{aligned}$$

a contradiction. A symmetric argument gives us that if $x_i(v, B') > x_i(v, B)$, there is some $\delta$ such that $u_i(v + \delta, B') < u_i(v + \delta, B)$.

On the other hand, if $u_i(v, B') > u_i(v, B)$, then we may conclude that $p_i(v, B') > B$ (since otherwise an agent with budget $B$ would have incentive to lie and report $B'$).

We can combine the above two observations to get the claimed characterization as follows. Consider the revenue curve for the maximum possible budget $\widehat{B}$, and for any other budget $B$. Then we know that at each $v$, either they are identical, or the payment for budget $\widehat{B}$ exceeds that for budget $B$. By continuity, the last point where they were identical must have had a corresponding price of $B$. After that point, the allocation curve for $B$ must remain constant, since any increase would imply an increase in price.

While the above assumed that $x_i(v, B)$ was continuous, note that the function is monotone and bounded. As such, it is continuous a.e., and so our characterization holds a.e. for general $x_i(v, B)$. □

We now give an example showing that when budgets are private the $\mathrm{E}[\sum_{i \in \mathcal{B}} B_i]$ benchmark cannot be approximated merely by mechanisms for $\widehat{\mathcal{I}}$.

Consider trying to serve a single agent with a fixed value for service of $v = 3n + 1$, and a budget $B$ distributed according to the distribution $G(B) = 1 - 1/B$ for $B \in [1, n)$ and $G(n) = 1$. If we consider the truncated value distribution, since we have a single agent offering a fixed price is optimal; but any fixed price $p$ gives revenue of $p(1 - F(p)) = 1$. On the other hand, consider offering the agent the menu

$$\{(1/4 + \alpha, 2\alpha^2 n) : \alpha \in [0, 3/4]\}.$$

Note that differentiating the agent's utility as a function of $\alpha$ gives us

$$\frac{d}{d\alpha}\left((1/4 + \alpha)(3n + 1) - 2\alpha^2 n\right) = 3n + 1 - 4\alpha n \geq 1$$

for $\alpha \in [0, 3/4]$, and so the agent always buys the most expensive lottery he or she can. Prices on the menu range from $0$ to $9n/8$, and each offer is made with probability at least $1/4$; so this menu extracts revenue of at least $\mathrm{E}[B]/4 = \Theta(\log n)$.

## B  Single agent revenue maximization

Here we present the missing proofs from Section 3. Recall that we are in a setting $\mathcal{I}$ with a single agent with value $v \sim F$, and public budget $B$. We first give the proofs of two claims regarding the revenue-optimal allocation rule $x^*$ for this setting.

*Proof of Claim 1.* Suppose that $x^*_{\max} < 1$, and consider setting $\hat{x}(v) = \frac{x^*(v)}{x^*_{\max}}$. Then we have that $\hat{x}_{\max} = 1$ and

$$\int (\hat{x}_{\max} - \hat{x}(v))dv = \frac{1}{x^*_{\max}} \int (x^*_{\max} - x^*(v))dv \leq B.$$

Furthermore $x^*$ must yield a nonnegative objective value (since $x(v) = 0$ is also valid solution). Thus, we get that

$$\int \hat{x}(v)\phi(v)f(v)dv \geq \int x^*(v)\phi(v)f(v)dv,$$

and so $\hat{x}$ only improves upon $x^*$. □

*Proof of Claim 2.* Suppose that $x^*(v^*) < 1$, and consider setting $\hat{x}(v) = x^*(v)$ for $v < v^*$ and $\hat{x}(v) = 1$ otherwise. Then we have

$$\int (1 - \hat{x}(v))dv \leq \int (1 - x^*(v))dv.$$

Furthermore,
$$\int \hat{x}(v)\phi(v)f(v)dv = \int_{v\leq v^*} x^*(v)\phi(v)f(v)dv + \int_{v\geq v^*} \phi(v)f(v)dv$$
$$\geq \int_{v\leq v^*} x^*(v)\phi(v)f(v)dv + \int_{v\geq v^*} x^*(v)\phi(v)f(v)dv$$
$$= \int x^*(v)\phi(v)f(v)dv,$$

since $\phi(v) \geq 0$ for $v \geq v^*$. This contradicts the optimality of $x^*$. $\square$

We now proceed to give the ironing procedure mentioned in Section 3. Recall that our goal is to find an optimal solution to
$$\min_{x\in\mathcal{A}} \int_0^{v^*} x(v)g(v)dv, \text{ where}$$
$$\mathcal{A} = \left\{\begin{array}{l} \text{increasing } x : [0, v^*] \to [0, 1] \\ \text{such that } \int_0^{v^*} x(v)dv = B' \end{array}\right\}.$$

The solution to the above is a simple step function if $g$ is non-increasing; however, if $g$ is not non-increasing, we need to "iron" the function $g$ to produce a non-increasing function $\hat{g}$. Let $G(v) = \int_0^v g(t)dt$; $\hat{G}(v)$ be the convex upper envelope of $G$; and $\hat{g} = \frac{d}{dv}\hat{G}(v)$. We will find it useful to focus on a subset of allocation functions that are compatible with the ironed $\hat{g}$ in that they are constant on ironed regions. Given any $v$ such that $G(v) \neq \hat{G}(v)$, define
$$\underline{v} = \sup\{v' \leq v : G(v') = \hat{G}(v')\}, \text{ and}$$
$$\bar{v} = \inf\{v' \geq v : G(v') = \hat{G}(v')\}.$$

Then $[\underline{v}, \bar{v}]$ is the ironed region containing $v$. Now, given any $x \in \mathcal{A}$, we define the modified allocation function $\tilde{x}$ by
$$\tilde{x}(v) = \begin{cases} x(v) & \text{if } G(v) = \hat{G}(v); \text{ and} \\ \frac{1}{\bar{v}-\underline{v}} \int_{\underline{v}}^{\bar{v}} x(t)dt & \text{if } G(v) \neq \hat{G}(v), \end{cases}$$

Note that $\tilde{x} \in \widetilde{\mathcal{A}}$.

We break the proof of Lemma 2 into the chain of comparisons
$$\int_0^{v^*} x(v)g(v)dv \geq \int_0^{v^*} x(v)\hat{g}(v)dv = \int_0^{v^*} \tilde{x}(v)\hat{g}(v)dv = \int_0^{v^*} \tilde{x}(v)g(v)dv.$$

We formalize each of these relations in a lemma; first, however, we state a fact that will be useful in their proofs.

**Fact 25.** *For any fixed $x \in \mathcal{A}$, and for any interval $(a, b)$ such that $G(v) = \hat{G}(v)$ for all $v \in (a, b)$, we have that*
$$\int_a^b x(v)g(v)dv = \int_a^b x(v)\hat{g}(v)dv.$$

We now proceed with proving the lemmas.

**Lemma 26.** *For any $x \in \mathcal{A}$, we have that $\int_0^{v^*} x(v)g(v)dv \geq \int_0^{v^*} x(v)\hat{g}(v)dv$.*

*Proof.* By Fact 25, we know we need only focus on regions where $\hat{g}$ is ironed. So let $v$ be such that $G(v) \neq \hat{G}(v)$, and consider the interval $[\underline{v}, \bar{v}]$.

By our definition of $\hat{G}(v)$, we know that for any $v' \in [\underline{v}, \bar{v}]$ we have
$$\int_{\underline{v}}^{v'} g(t)dt \leq \int_{\underline{v}}^{v'} \hat{g}(t)dt,$$

with equality if and only if $v' = \underline{v}$ or $v' = \bar{v}$. This implies the following. Let

$$\gamma = \int_{\underline{v}}^{\bar{v}} g(t)dt = \int_{\underline{v}}^{\bar{v}} \hat{g}(t)dt;$$

then both $G(\cdot)/\gamma$ and $\hat{G}(\cdot)/\gamma$ are distributions on $(\underline{v}, \bar{v})$, and the former stochastically dominates the latter. As such, since $x(\cdot)$ is monotone increasing, we must have that the expectations of $x(\cdot)$ under these two distributions are related as

$$\int_{\underline{v}}^{\bar{v}} x(t)\frac{g(t)}{\gamma}dt \geq \int_{\underline{v}}^{\bar{v}} x(t)\frac{\hat{g}(t)}{\gamma}dt \iff \int_{\underline{v}}^{\bar{v}} x(t)g(t)dt \geq \int_{\underline{v}}^{\bar{v}} x(t)\hat{g}(t)dt.$$

Since the claimed relation holds on ironed intervals as well, we may conclude that it holds overall. □

**Lemma 27.** *For any $x \in \mathcal{A}$, we have that $\int_0^{v^*} x(v)\hat{g}(v)dv = \int_0^{v^*} \tilde{x}(v)\hat{g}(v)dv$.*

*Proof.* First, note that by our definition of $\tilde{x}$, we know that $\tilde{x}(v) = x(v)$ whenever $G(v) = \hat{G}(v)$. This means we once again need only consider ironed regions. Let $v$ be such that $G(v) \neq \hat{G}(v)$, and consider the interval $[\underline{v}, \bar{v}]$. Now, we know that on $(\underline{v}, \bar{v})$, $\hat{g}(\cdot)$ takes on the constant value $\hat{g}(v)$. So we have that

$$\int_{\underline{v}}^{\bar{v}} x(t)\hat{g}(t)dt = \hat{g}(v)\int_{\underline{v}}^{\bar{v}} x(t)dt = \hat{g}(v)(\bar{v} - \underline{v})\tilde{x}(t) = \int_{\underline{v}}^{\bar{v}} \tilde{x}(t)\hat{g}(t)dt,$$

and the claim follows. □

**Lemma 28.** *For any $\tilde{x} \in \widetilde{\mathcal{A}}$, we have that $\int_0^{v^*} \tilde{x}(v)\hat{g}(v)dv = \int_0^{v^*} \tilde{x}(v)g(v)dv$.*

*Proof.* Once again, Fact 25 implies we know we need only focus on regions where $\hat{g}$ is ironed. So let $v$ be such that $G(v) \neq \hat{G}(v)$, and consider the interval $[\underline{v}, \bar{v}]$. Then we know that $\tilde{x}(\cdot)$ is constant on the interval $(\underline{v}, \bar{v})$. Thus,

$$\int_{\underline{v}}^{\bar{v}} \tilde{x}(t)\hat{g}(t)dt = \tilde{x}(v)\int_{\underline{v}}^{\bar{v}} \hat{g}(t)dt = \int_{\underline{v}}^{\bar{v}} \tilde{x}(t)g(t)dt,$$

and the result follows. □

## C Revenue maximization with private budgets

Let $\mathcal{I} = (\mathbf{F}, \mathcal{S}, \mathbf{G})$ denote an instance of this setting. In Section 3, we presented a mechanism that was approximately optimal under the condition that every distribution in $\mathbf{F}$ satisfied the MHR condition. Here, show how we can modify the mechanism to address settings without an MHR assumption, though we still need a technical condition on values and budgets.

Appendix A provides a characterization of truthful mechanisms for the setting $\mathcal{I}$. Given this characterization, we proceed with a reduction similar to our approach for public budgets. Specifically, given a mechanism $M$ for the setting $\mathcal{I}$, our goal is to use (3) as an upper bound on revenue. We define the mechanism $\widehat{M}$ for the single-parameter setting with feasibility constraint $\mathcal{S}$ and value distribution $\widehat{\mathbf{F}}$ as follows. Let $\mathbf{x}(\mathbf{v}, \mathbf{B})$ and $\mathbf{p}(\mathbf{v}, \mathbf{B})$ be the allocation and payment rules for $M$; define the expected allocation and payment for $\widehat{M}$ as follows. For each agent $i$ with valuation $\hat{v}_i$, draw a corresponding pair $(v_i, B_i)$ consistent with $\hat{v}_i = \min(v_i, B_i)$; $\widehat{M}$ has expected allocation and payment

$$\hat{x}_i(\hat{\mathbf{v}}) = x_i((\mathbf{v}_{-i}, \hat{v}_i), (\mathbf{B}_{-i}, \hat{v}_i)); \text{ and}$$
$$\hat{p}_i(\hat{\mathbf{v}}) = p_i((\mathbf{v}_{-i}, \hat{v}_i), (\mathbf{B}_{-i}, \hat{v}_i)),$$

respectively. Noting that $x_i(\mathbf{v}, \mathbf{B})$ must be monotone in both $v_i$ and $B_i$, and hence $\hat{x}_i(\hat{\mathbf{v}}) \leq x(\mathbf{v}, \mathbf{B})$, and following the proof of Lemma 4 we conclude once again that $\widehat{M}$ is a feasible BIC mechanism for $\widehat{\mathbf{F}}$. Also note that since EPIR implies that $p_i(\mathbf{v}, \mathbf{B}) \leq v_i$, our characterization of truthfulness implies that $p_i(\mathbf{v}, (\mathbf{B}_{-i}, B_i)) = p_i(\mathbf{v}, (\mathbf{B}_{-i}, B_i'))$ for any $B_i, B_i' \geq v_i$.

The preceding observations on $\widehat{M}$, along with the fact that no agent ever pays more than their budget, immediately imply the following lemma.

**Lemma 29.** *For $\widehat{M}$ as defined above, we have that $\mathcal{R}^M \leq \mathcal{R}^{\widehat{M}} + \mathrm{E}[\sum_{i \in \mathcal{B}} B_i]$.*

$\mathcal{R}^{\widehat{M}}$ is once again dominated by the revenue of an optimal mechanism over the instance $\widehat{\mathcal{I}}$. Bounding $\mathrm{E}[\sum_{i \in \mathcal{B}} B_i]$ is quite tricky, however. In particular, we show in Appendix A that this benchmark cannot be approximated by a mechanism for $\widehat{\mathcal{I}}$. The example in the appendix motivates using menus over lotteries (or randomized allocations) so that a rational agent maximizes her utility by picking options with payments as close to her budget as possible. We will now define a mechanism $M_{\mathcal{B}}$ for the original setting $\mathcal{I}$ based on lottery menus that allows us to complete our reduction. We begin by defining a class of lottery menus similar to those used in Section 3, and proving some revenue properties about them; we then continue on to show how we use these offers to approximate our stated benchmark.

Given a maximum budget $\widehat{B}$, and a threshold $T$, we define a lottery system $\mathcal{L}(\widehat{B}, T)$ as follows. The elements of $\mathcal{L}(\widehat{B}, T)$ are parametrized by a value $\alpha \in [0, 1/2]$, and have the form $\ell_\alpha = (1/2 + \alpha, \alpha^2 \widehat{B}/2)$, which we interpret as, with probability $(1/2 + \alpha)$, offering the agent a chance to buy the item at price

$$\frac{\widehat{B}}{2} \cdot \alpha^2 \cdot (1/2 + \alpha)^{-1} = \widehat{B}\left(\frac{\alpha^2}{1 + 2\alpha}\right).$$

The lottery system $\mathcal{L}(\widehat{B}, T)$ consists of all $\ell_\alpha$ which ask a price of at least $T$. Let $\hat{\alpha}$ denote the minimum such $\alpha$. If it is the case that $\hat{\alpha} > 1/2$ (and so the above set is empty), we explicitly define $\mathcal{L}(\widehat{B}, T)$ to consist of a single lottery that always offers the item at a price of $T$, i.e. $\mathcal{L}(\widehat{B}, T) = \{(1, T)\}$.

Consider offering the lottery system $\mathcal{L}(\widehat{B}, T)$ to an agent with valuation $v$ and budget $B \in [0, \widehat{B}]$. From our definition of the lottery system $\mathcal{L}(\widehat{B}, T)$, we get the following lemma.

**Lemma 30.** *Given that $v, B \geq T$, there is at least one lottery in $\mathcal{L}(\widehat{B}, T)$ that provides the agent non-negative utility, and the system obtains revenue of at least $T/2$. Furthermore, if $v \geq B$, the revenue obtained is at least*

$$B/8 \cdot \min(1, 4v/\widehat{B}).$$

*Proof.* We begin by considering the case where $\mathcal{L}(\widehat{B}, T)$ contains the single offer $(1, T)$. Note that in this case, every possible $\ell_\alpha$ had a price below $T$; in particular, this holds for the price corresponding to $\alpha = 1/2$, which is $\widehat{B}/8$. Since $v, B \geq T$ by assumption, the lottery offered provides non-negative utility, and the claim follows immediately. So we assume we are not in this case for the rest of the proof.

Note the following about the lottery system $\mathcal{L}(\widehat{B}, T)$.

- Since we always provide a lottery that allows the agent, with some probability, to purchase the item at a price of $T$, and by assumption $v, B \geq T$, there is always at least one option in the lottery system providing non-negative utility to the agent.

- Every option in $\mathcal{L}(\widehat{B}, T)$ results in the agent buying the item with probability at least $1/2$, and so revenue is always at least $1/2$ of the price corresponding to the option they choose. Note that this immediately implies that expected revenue is always at least $T/2$, since we exclude any options with price below $T$

- Given that an agent with value $v$ selects some $\ell_\alpha$, their utility as a function of $\alpha$ is

$$u(\alpha) = (1/2 + \alpha)\left(v - \widehat{B}\left(\frac{\alpha^2}{1 + 2\alpha}\right)\right)$$

$$= (1/2 + \alpha)v - \frac{\alpha^2 \widehat{B}}{2},$$

and so

$$u'(\alpha) = v - \alpha\widehat{B}, \text{ and}$$
$$u''(\alpha) = -\widehat{B}.$$

So we can see that utility as a function of $\alpha$ is a concave function with maximum at $\alpha^* = v/\widehat{B}$, which corresponds to a lottery $\ell_{\alpha^*}$ with expected revenue of $v^2/2\widehat{B}$. It may not always be the case, however, that $\alpha^*$ corresponds to a lottery in the system $\mathcal{L}(\widehat{B}, T)$. Since utility is a concave function of $\alpha$, we can see that if $\alpha^*$ is below or above the range of $\alpha$ included in $\mathcal{L}(\widehat{B}, T)$, the agent purchases the lowest or highest $\alpha$ that *is* included, respectively. In the former case, we only do better than $v^2/2\widehat{B}$, since by inspection we can see that expected revenue from $\ell_\alpha$ is an increasing function of $\alpha$, and in the latter our expected revenue is precisely $\widehat{B}/8$. So in any case, our revenue is at least

$$\min(v^2/2\widehat{B}, \widehat{B}/8) \geq \frac{B}{8} \min(1, 4v/\widehat{B}),$$

if we assume that $v \geq B$.

Thus, in either case we get that revenue satisfies the claimed bound. $\square$

To bound $\mathrm{E}[\sum_{i \in \mathcal{B}} B_i]$, our mechanism serves the set $\mathcal{A}$ defined by $\mathcal{A}(\mathbf{v}, \mathbf{B}) = \mathrm{argmax}_{S \in \mathcal{S}} \sum_{i \in S} \min(v_i, B_i)$. For each $i$, let $T_i(\mathbf{v}, \mathbf{B})$ be the truthful price corresponding to the allocation $\mathcal{A}$, i.e. $T_i(\mathbf{v}, \mathbf{B}) = \min\{v' | i \in \mathcal{A}((\mathbf{v}_{-i}, v'), \mathbf{B})\}$. Our mechanism offers agent $i$ the lottery system $\mathcal{L}(\widehat{B}_i, T_i)$.

**Lemma 31.** *If $\mathcal{S}$ is a matroid constraint, and for all $i$ we have that $\Pr[v_i \geq \widehat{B}_i/4] \geq 1/2$, we can construct a mechanism $M$ whose expected revenue 16-approximates $\mathrm{E}[\sum_{i \in \mathcal{B}} B_i]$.*

*Proof.* We show that as long as $\mathcal{S}$ is a matroid constraint, we can actually achieve revenue that approximates our benchmark $\mathrm{E}[\sum_{i \in \mathcal{B}} B_i]$.

We begin by noting that while it would be natural to try and serve the set $\mathcal{B}$ using lottery systems $\mathcal{L}(\widehat{B}, T)$ as described previously, this is not truthful; in particular, membership in $\mathcal{B}$ is not monotone in $B_i$.

Instead, we serve the set $\mathcal{A}$ defined by

$$\mathcal{A}(\mathbf{v}, \mathbf{B}) = \mathrm{argmax}_{S \in \mathcal{S}} \sum_{i \in S} \min(v_i, B_i),$$

and show that we can only do better than we would have by serving $\mathcal{B}$. For each $i$, let $T_i(\mathbf{v}, \mathbf{B})$ be the truthful price corresponding to the allocation $\mathcal{A}$, i.e.

$$T_i(\mathbf{v}, \mathbf{B}) = \min\{v' | i \in \mathcal{A}((\mathbf{v}_{-i}, v'), \mathbf{B})\}.$$

Our mechanism offers agent $i$ the lottery system $\mathcal{L}(\widehat{B}_i, T_i)$; if none of the options in $\mathcal{L}(\widehat{B}_i, T_i)$ is both desirable and feasible for the agent, they may choose to buy nothing instead. Recall that every option in the lottery system $\mathcal{L}(\widehat{B}_i, T_i)$ charges the agent a price of at least $T_i$; combining this with Lemma 30, we can see that each agent $i$ chooses an option from their menu if and only if $i \in \mathcal{A}$. Since $\mathcal{A} \in \mathcal{S}$, this implies the mechanism makes a feasible allocation.

Consider fixing the sets $\mathcal{B}$ and $\mathcal{A}$. Note that since $\mathcal{S}$ is a matroid, $\mathcal{A}$ is always maximal; in particular, this means that if $\mathcal{A} \cup \{i\} \in \mathcal{S}$, then $i \in \mathcal{A}$. Furthermore, since $\mathcal{B}$ is maximal as well, for every pair of vectors $(\mathbf{v}, \mathbf{B})$, we can get a 1-1 function $g : \mathcal{B} \to \mathcal{A}$ such that for all $i \in \mathcal{B}$, we have $\mathcal{B} \setminus \{g(i)\} \cup \{i\} \in \mathcal{S}$. Note that by the definition of $g$ and $T_i(\mathbf{v}, \mathbf{B})$, one of the following must hold for each $i \in \mathcal{B}$.

- $i \in \mathcal{A}$ as well. Note that $i \in \mathcal{A}$ implies $v_i, B_i \geq T_i(\mathbf{v}, \mathbf{B})$, and $i \in \mathcal{B}$ implies $v_i \geq B_i$. So we can apply Lemma 30 to see that

$$\mathop{\mathrm{E}}_{\mathbf{v}, \mathbf{B}}\left[\min(1, 4v_i/\widehat{B}_i)\Big|\mathcal{A}, \mathcal{B}\right] = \mathop{\mathrm{E}}_{\mathbf{v}_{-i}, \mathbf{B}}\left[\mathop{\mathrm{E}}_{v_i}\left[\min(1, 4v_i/\widehat{B}_i)\Big|\mathcal{A}, \mathcal{B}, \mathbf{v}_{-i}, \mathbf{B}\right]\Big|\mathcal{A}, \mathcal{B}\right]$$

$$\geq \mathop{\mathrm{E}}_{\mathbf{v}_{-i}, \mathbf{B}}\left[\mathop{\mathrm{E}}_{v_i}\left[\min(1, 4v_i/\widehat{B}_i)\right]\Big|\mathcal{A}, \mathcal{B}\right] \quad (4)$$

$$\geq \mathop{\mathrm{E}}_{\mathbf{v}_{-i}, \mathbf{B}}\left[1/2|\mathcal{A}, \mathcal{B}\right],$$

and so

$$
\begin{align}
\mathop{\mathrm{E}}_{\mathbf{v},\mathbf{B}}\left[\mathcal{R}_i^M(\mathbf{v},\mathbf{B})\big|\mathcal{A},\mathcal{B}\right] &\geq \mathop{\mathrm{E}}_{\mathbf{v},\mathbf{B}}\left[B_i/8 \cdot \min(1, 4v_i/\widehat{B}_i)\big|\mathcal{A},\mathcal{B}\right] \\
&\geq \mathop{\mathrm{E}}_{\mathbf{v},\mathbf{B}}[B_i/8|\mathcal{A},\mathcal{B}] \cdot \mathop{\mathrm{E}}_{\mathbf{v}_{-i},\mathbf{B}}\left[\mathop{\mathrm{E}}_{v_i}\left[\min(1,4v_i/\widehat{B}_i)\right]\bigg|\mathcal{A},\mathcal{B}\right] \tag{5} \\
&\geq \frac{1}{8} \cdot \mathop{\mathrm{E}}_{\mathbf{v},\mathbf{B}}[B_i|\mathcal{A},\mathcal{B}] \cdot \mathop{\mathrm{E}}_{\mathbf{v}_{-i},\mathbf{B}}[1/2|\mathcal{A},\mathcal{B}] \\
&\geq \frac{1}{16} \cdot \mathop{\mathrm{E}}_{\mathbf{v},\mathbf{B}}[B_i|\mathcal{A},\mathcal{B}],
\end{align}
$$

where (5) follows from the independence of $v_i$ and $B_i$, and (4) follows from the fact that the conditioning in the inner expectation translates into a lower bound on $v_i$, which can only increase the expectation.

- In the other case, we have that $i \notin \mathcal{A}$, but $\mathcal{A} \setminus \{g(i)\} \cup \{i\} \in \mathcal{S}$, and so by the definition of $T(\mathbf{v},\mathbf{B})$ we must have that $T_{g(i)}(\mathbf{v},\mathbf{B}) \geq B$. So we can once again apply Lemma 30 to see that

$$
\mathop{\mathrm{E}}_{\mathbf{v},\mathbf{B}}\left[\mathcal{R}_{g(i)}^M(\mathbf{v},\mathbf{B})\big|\mathcal{A},\mathcal{B}\right] \geq \mathop{\mathrm{E}}_{\mathbf{v},\mathbf{B}}\left[T_{g(i)}(\mathbf{v},\mathbf{B})/2\big|\mathcal{A},\mathcal{B}\right] \geq 1/2 \cdot \mathop{\mathrm{E}}_{\mathbf{v},\mathbf{B}}[B_i|\mathcal{A},\mathcal{B}].
$$

Combining the above two arguments, we can see that for any fixed $\mathcal{A}$ and $\mathcal{B}$ we have that

$$
\begin{align}
\mathop{\mathrm{E}}_{\mathbf{v},\mathbf{B}}\left[\sum_{i\in\mathcal{B}} B_i \bigg| \mathcal{A},\mathcal{B}\right] &\leq 16 \mathop{\mathrm{E}}_{\mathbf{v},\mathbf{B}}\left[\sum_{i\in\mathcal{B}\cap\mathcal{A}} \mathcal{R}_i^M(\mathbf{v},\mathbf{B}) \sum_{i\in\mathcal{B}\setminus\mathcal{A}} \mathcal{R}_{g(i)}^M(\mathbf{v},\mathbf{B}) \bigg| \mathcal{A},\mathcal{B}\right] \\
&\leq 16 \mathop{\mathrm{E}}_{\mathbf{v},\mathbf{B}}\left[\sum_{i\in\mathcal{A}} \mathcal{R}_i^M(\mathbf{v},\mathbf{B}) \bigg| \mathcal{A},\mathcal{B}\right] \\
&= 16 \mathop{\mathrm{E}}_{\mathbf{v},\mathbf{B}}\left[\mathcal{R}^M(\mathbf{v},\mathbf{B})\big|\mathcal{A},\mathcal{B}\right];
\end{align}
$$

taking expectations over $\mathcal{A}$ and $\mathcal{B}$ completes the argument. □

We obtain the following result:

**Theorem 32.** *Suppose that every agent's median value is no smaller than a constant fraction of her maximum budget. Then we can construct a budget-feasible mechanism that is DSIC with respect to both values and budgets, and obtains a constant fraction of the revenue of the optimal such mechanism.*